\documentclass[aps,prc,groupedaddress,nofootinbib,twocolumn]{revtex4}

\usepackage[dvips]{graphicx}
\usepackage{amssymb,amsmath,amstext,amsthm,amsfonts}
\usepackage{slashed}

\usepackage[usenames]{color}
\usepackage{pstricks}

\newcommand{\lep}{\ell}
\newcommand{\kpr}{k'}
\newcommand{\myvec}[1]{\mathbf{#1}}
\newcommand{\absp}{|\myvec{p}|}
\newcommand{\abskpr}{|\myvec{\kpr}|}
\newcommand{\res}[3]{#1$_{#2}$(#3)}
\newcommand{\dsdqs}{\dd \sigma / \dd Q^2}

\newcommand{\reffig}[1]{Fig.~\ref{#1}}
\newcommand{\refeq}[1]{Eq.~(\ref{#1})}
\newcommand{\reftab}[1]{Table~\ref{#1}}
\newcommand{\refsec}[1]{Sec.~\ref{#1}}
\newcommand{\refetal}[1]{\emph{et~al.}\ \cite{#1}}
\newcommand{\refcite}[1]{Ref.~\cite{#1}}

\newcommand{\atom}[2]{\mbox{$^{#1}\text{#2}$}}
\newcommand{\carbon}{{\atom{12}{C}}}
\newcommand{\plotscale}{.70}
\newcommand{\myunit}[1]{\mbox{$\,\text{#1}$}}
\newcommand{\GeV}{\myunit{GeV}}
\newcommand{\MeV}{\myunit{MeV}}

\newcommand{\dd}{\mathrm{d}}

\begin{document}

\title{Neutrino-nucleus scattering reexamined: Quasielastic scattering and pion production
  entanglement and implications for neutrino energy reconstruction}

\author{T.~Leitner}
\email{leitner@theo.physik.uni-giessen.de}
\affiliation{Institut f\"ur Theoretische Physik, Universit\"at Giessen, Germany}

\author{U.~Mosel}
\affiliation{Institut f\"ur Theoretische Physik, Universit\"at Giessen, Germany}

\date{23 April 2010}

\begin{abstract}

  We apply the GiBUU model to questions relevant for current and future long-baseline
  neutrino experiments, we address in particular the relevance of charged-current
  reactions for neutrino-disappearance experiments. A correct identification of
  charged-current quasielastic (CCQE) events --- which is the signal channel in
  oscillation experiments --- is relevant for the neutrino energy reconstruction and thus
  for the oscillation result. We show that about 20\% of the quasielastic cross section is
  misidentified in present-day experiments and has to be corrected for by means of event
  generators. Furthermore, we show that a significant part of $1\pi^+$ ($>$ 40\%) events
  is misidentified as CCQE events mainly caused by pion absorption in the nucleus.  We
  also discuss the dependence of both of these numbers on experimental detection
  thresholds. We further investigate the influence of final-state interactions on the
  neutrino energy reconstruction.

\end{abstract}


\maketitle


\section{Introduction}

Good knowledge of the neutrino energy is required for a precise determination of
oscillation parameters in $\nu_\mu$ disappearance measurements. However, the neutrino beam
is far from being mono-energetic in present experiments. Thus, $\nu_\mu$ disappearance
experiments search for a distortion in the neutrino flux in the detector positioned far
away from the source. Comparing both, un-oscillated and oscillated flux, one gains
information about the oscillation probability and with that about mixing angles and mass
squared differences.

The neutrino energy is not measurable directly but has to be reconstructed from the
reaction products.\footnote{ External experiments are performed to help predict the beam
  profile. See, for example, the hadron production experiment at CERN (HARP)
  \cite{HARPWebsite}.} Present oscillation experiments use the charged-current
quasielastic (CCQE) reaction as signal event and reconstruct the energy with quasifree
two-body kinematics from the outgoing muon assuming the target nucleon is at rest.  Two
immediate questions arise from this procedure: (1) How good is the identification of CCQE
events?  (2) How exact is the assumption of quasifree two-body kinematics for nucleons
bound in a nucleus where many in-medium modifications are present?

CCQE is defined as $\nu n \to \lep^- p$ (i.e., on the nucleon). In the nucleus, CCQE is
masked by final-state interactions (FSI). Thus, the correct identification of CCQE is
immediately related to the question of how FSI influence the event selection. The main
background to CCQE is CC$1\pi^+$ production. If the pion is absorbed in the nucleus and/or
not seen in the detector, these events can be misidentified as CCQE events. Consequently,
a proper understanding of both CCQE \emph{and} CC$1\pi^+$ on nuclei is essential for the
reconstruction of the neutrino energy.

This article addresses the questions outlined above in the framework of the GiBUU
transport model. After giving the necessary model details, we start with a general
introduction of event classification in typical neutrino detectors and discuss how
possible detection thresholds influence the measured spectra.  We further discuss both
CCQE and CC$1\pi^+$ cross sections and their entanglement. Finally, we investigate how
nuclear effects influence the reconstruction of the neutrino energy.


\section{Neutrino scattering in the GiBUU transport model}

The presence of in-medium modifications and, in particular, final-state interactions
inside the target nucleus requires the use of state-of-the-art theoretical methods for the
extraction of elementary processes from experiments with nuclear targets (see also the
review by Alvarez-Ruso \cite{AlvarezRuso:2009mn}). For this aim we use the GiBUU transport
model \cite{gibuu}, where the neutrino first interacts with a bound nucleon. It has
recently been questioned how good this impulse approximation (IA) actually is
\cite{Ankowski:2010yh,Martini:2009uj}. Our experience with photonuclear processes
\cite{Krusche:2004uw} indicates that this approximation gives reasonably reliable results
for momentum transfers $>$ 0.2 GeV.  The use of the IA requires a good description of both
the elementary vertex and the in-medium-modifications. The final state of this initial
reaction undergoes complex hadronic final-state interactions. In the following, we give a
brief overview of our model and refer the reader in particular to \refcite{Leitner:2008ue}
and to our website \cite{gibuu} for more details.

In the energy region of $E_\nu \sim 0.5-2 \GeV$, the elementary neutrino-nucleon cross
section contains contributions from quasielastic scattering (QE: $ \ell N \to \ell' N'$),
resonance excitation (R: $ \ell N \to \ell' R$) and direct (i.e., nonresonant)
single-pion production (BG: $ \ell N \to \ell' \pi N'$) treated in our description as
background. QE scattering is the most important process at these energies, followed by
single-pion production through the excitation and subsequent decay of the $\Delta$
resonance \res{P}{33}{1232}. However, we include in addition 12 $N^*$ and $\Delta$
resonances with invariant masses less than 2 GeV. The vector parts of the single
contributions are obtained from recent analyses of electron-scattering cross sections.
The axial couplings are obtained from PCAC (partial conservation of the axial current),
and, wherever possible, we use neutrino-nucleon scattering data as input.

The elementary neutrino-nucleon cross section is modified in the nuclear medium. Bound
nucleons are treated within a local Thomas-Fermi approximation which naturally includes
Pauli blocking. The nucleons are bound in a mean-field potential depending on density and
momentum which we account for by evaluating the above cross sections with full in-medium
kinematics.  We further consider the collisional broadening of the final-state particles
within the low-density approximation $\Gamma_\text{coll} = \rho \sigma v$ obtained in a
consistent way from the GiBUU cross sections.  Details of our model for the elementary
vertex and the corresponding medium modifications can be found in
\refcite{Leitner:2008ue}.

In the next step, the particles propagate through the nucleus undergoing final-state
interactions (FSI) which are simulated with the coupled-channel semiclassical GiBUU
transport model (more information and code download on our website~\cite{gibuu}).  The
GiBUU model is based on well-founded theoretical ingredients and has been tested in
various very different nuclear reactions; in particular, against electron- and
photon-scattering data \cite{Krusche:2004uw,Leitner:2008ue}.

The space-time evolution of a many-particle system in a mean-field potential is described
by the BUU equation. For particles of species $i$, it is given by
\begin{multline}
  \left({\partial_t}+\myvec\nabla_p H\cdot\myvec\nabla_r -\myvec\nabla_r
    H\cdot\myvec\nabla_p\right)f_i(\myvec{r},p, t) = \\ I_\text{coll}
  [f_i,f_N,f_\pi,f_{\Delta},...],
\end{multline}
where the phase-space density $f_i(\myvec{r},p,t)$ depends on time $t$, coordinates
$\myvec{r}$ and the four-momentum $p$.  $H$ is the relativistic Hamiltonian of a particle
of mass $M$ in a scalar potential $U$ given by $H=\left(\left[ M + U(\myvec{r},p
    )\right]^2 + \myvec{p}^{\,2} \right)^{1/2}$. The scalar potential $U$ usually depends
both on four momentum and on the nuclear density.
The BUU equations are coupled through the collision term $I_\text{coll}$ which accounts
for changes (gain and loss) in the phase-space density due to elastic and inelastic
collisions between particles and also to particles decaying into other hadrons. In
particular, we include two-body reactions like e.g.\ $\pi N \to \pi N$, $NN \to NN$, $R N
\to N N$, $R N \to R' N$, and three-body processes like $\pi N N \to NN$ and $\Delta N N
\to NNN$. By this coupled-channel treatment we can describe side-feeding processes into
different channels. This complex set of coupled differential-integral equations is then
solved numerically with the GiBUU code.

All particles (also resonances) are propagated in mean-field potentials according to the
BUU equations. Those states acquire medium-modified spectral functions (nucleons and
resonances) and are propagated off shell.  The medium modification of the spectral
function is based both on collisional broadening and on the mean-field potentials, both of
which depend on particle kinematics as well as on nuclear density.

Altogether, FSI lead to absorption, charge exchange and redistribution of energy and
momentum, as well as to the production of new particles. We have shown in earlier works,
that their impact on neutrino-induced pion production is dramatic
\cite{Leitner:2006ww,Leitner:2008wx}. Thus, a qualitatively and quantitatively correct
treatment of these effects is of great importance, especially for the energy
reconstruction as we will demonstrate in the following.


\section{Event selection}
\label{sec:eventselection}

Event selection in current neutrino experiments is a highly complicated subject.  Rather
than presenting a quantitative discussion for each particular setup we give a qualitative
picture of how nuclear effects themselves modify the measured spectra in charged-current
(CC) scattering assuming certain detection methods. Two generic detectors with the
following properties are used toward this aim. We note that the event identifications used
here are the ones used in the actual experiments.
\begin{description}
\item[Cherenkov detector.] In a Cherenkov detector (e.g., MiniBooNE and K2K-1kt), CCQE
  events are identified by a single ring from the outgoing lepton. Muons can be tagged by
  their decay electron. If pions are produced, they lead to additional rings either from
  the $\gamma$ decay of the $\pi^0$ or from the decay muon of the charged pions.

  For the Cherenkov detector, we identify the two relevant processes in the following way:

  \begin{tabular}{l l l l l l l}
    CCQE:       &$  1 \mu^- $&$ 0 \pi^+ $&$ 0 \pi^- $&$ 0 \pi^0 $&$ x p $&$ x n $ , \\
    CC$1\pi^+$: &$  1 \mu^- $&$ 1 \pi^+ $&$ 0 \pi^- $&$ 0 \pi^0 $&$ x p $&$ x n $ ,  
  \end{tabular}

  where $x p$ and $x n$ indicate, that any number of protons or neutrons are allowed.

  The lower momentum thresholds depend on the index of refraction $n$ via
  \begin{equation}
    \beta_\text{thres}=\frac1n \Leftrightarrow \absp_\text{thres} = \frac{m}{\sqrt{n^2 - 1}} \; ,
  \end{equation}
  where $m$ is the particle mass. From this, one easily obtains the kinetic energy
  thresholds. Typical values for water ($n=1.33$) are $T_\text{thres}\approx$ 55 MeV for
  muons, 75 MeV for charged pions, 0 MeV for neutral pions (identified via their $\gamma$
  decay), and 485 MeV for protons.  Lower thresholds ($\approx 10$ MeV for muons and
  charged pions, $\approx 65$ MeV for protons) are reached with the MiniBooNE detector
  which is filled with mineral oil with $n=1.47$ \cite{AguilarArevalo:2008qa} and, in
  addition, produces scintillation light.

\item[Tracking detector.] In a tracking detector (e.g., SciBooNE and K2K SciFi), all
  charged particles leave tracks which can be used to identify the particles and determine
  their properties. Thus, highly advanced event selection procedures are applied. To keep
  it simple, we identify

  \begin{tabular}{l l l l l l l}
    CCQE:       &$  1 \mu^- $&$ 0 \pi^+ $&$ 0 \pi^- $&$ 0 \pi^0 $&$ 1 p $&$ x n $ , \\
    CC$1\pi^+$: &$  1 \mu^- $&$ 1 \pi^+ $&$ 0 \pi^- $&$ 0 \pi^0 $&$ x p $&$ x n $ .  
  \end{tabular}

  The thresholds depend strongly on the experimental setup, e.g., the SciFi detector
  requires both muon (pion) kinetic energy to be above $\approx$ 500 MeV (100 MeV) and the
  proton kinetic energy above 175 MeV \cite{Gran:2006jn}.
 
\end{description}

In the following, we assume perfect particle identification above threshold in both cases
and neglect any other experimental restrictions.


\section{Topologies}
\label{sec:topologies}

\subsection{CCQE identification}

\begin{figure}[tbp]
  \centering
  \includegraphics[scale=\plotscale]{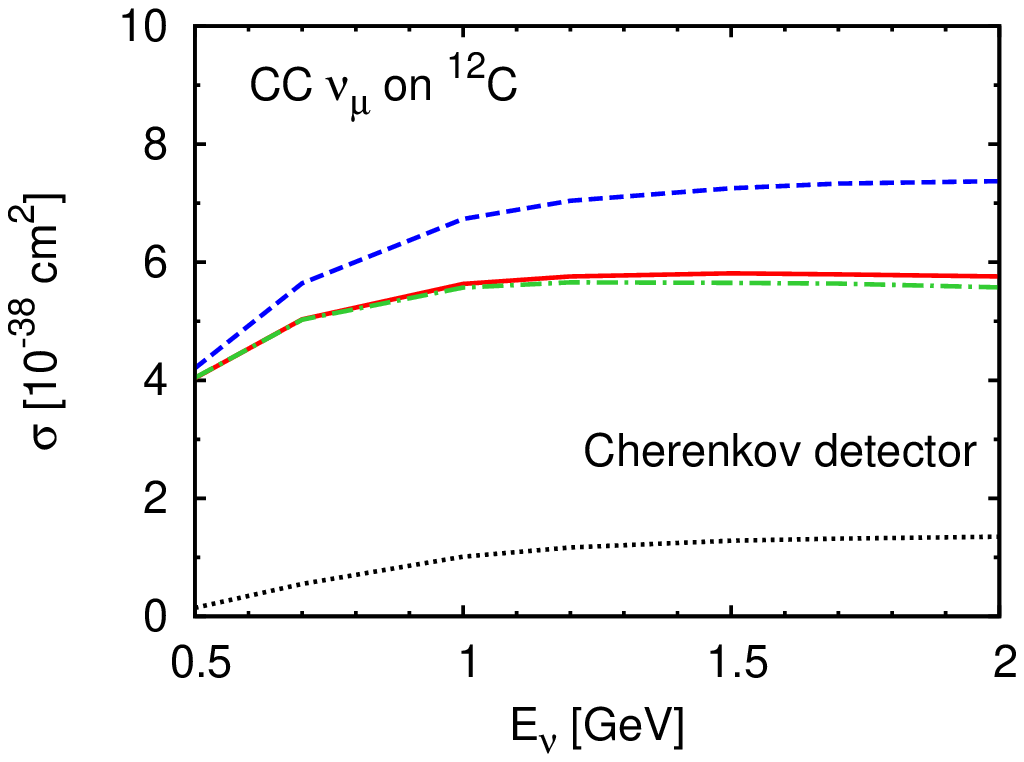}
  \includegraphics[scale=\plotscale]{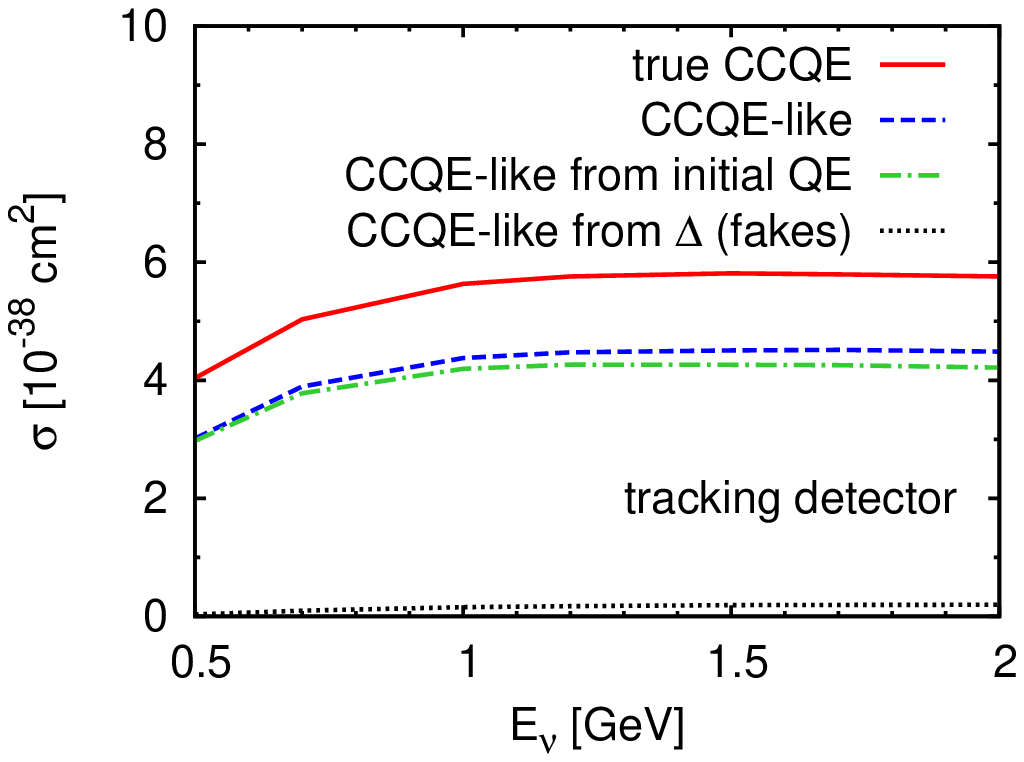}
  \caption{(Color online) Total QE cross section on \carbon{} (solid lines) compared to
    different methods on how to identify CCQE-like events in experiments (dashed lines).
    The top panel shows the method commonly applied in Cherenkov detectors; the lower
    panel shows the tracking-detector method as described in the text. The contributions to the
    CCQE-like events are also classified [CCQE-like from initial QE (dash-dotted) and from
    initial $\Delta$ (dotted lines)]. Experimental detection thresholds are not taken into
    account.
    \label{fig:QEmethods}}
\end{figure}

The CCQE reaction, $\nu_\lep n \to \lep^- p$, being the dominant cross section at low
energies, is commonly used to reconstruct the neutrino energy. In other words, CCQE is the
signal event in the present oscillation experiments.

The experimental challenge is to identify \emph{true} CCQE events in the detector, namely,
muons originating from an initial QE process. To be more precise, true CCQE corresponds to
the inclusive CCQE cross section including all medium effects or, in other words, the
CCQE cross section before FSI.  The difficulty comes from the fact that the true CCQE
events are masked by FSI in a detector built from nuclei.  The FSI lead to misidentified
events, e.g., an initial $\Delta$ whose decay pion is absorbed or which undergoes
``pion-less decay'' contributes to knock-out nucleons and can thus be counted as a CCQE
event --- we call this type of background event a ``fake CCQE'' event. We denote every
event which looks like a CCQE event as ``CCQE-like''.

As outlined above, in Cherenkov detectors CCQE-like events are all those where no pion is
detected, whereas in tracking detectors, CCQE-like events are those where a single proton
track is visible and at the same time no pions are detected.  The two methods are compared
in \reffig{fig:QEmethods}. The ``true CCQE'' events are denoted with solid lines, the
CCQE-like events by dashed lines. The Cherenkov detector is able to detect almost all true
CCQE events (top panel; solid vs.\ dash-dotted lines approximately agree) but sees also a
considerable amount of ``fake CCQE'' (or ``non-CCQE'') events (top panel; the dashed line
is roughly 20\% higher than the solid line). They are caused mainly by initial $\Delta$
excitation as described in the previous paragraph (absorption of decay pion or ``pion-less
decay''); their contribution to the cross section is given by the dotted lines.  These
additional (fake) events have to be removed from the measured event rates by means of
event generators, if one is interested only in the true QE events. It is obvious that this
removal is better the more realistic the generator is in handling the in-medium $\pi N
\Delta$ dynamics. On the contrary, less CCQE-like than true CCQE events are detected using
the method applied in tracking detectors, which triggers both on pions and protons (lower
panel, difference between dashed and solid line).  The FSI of the initial proton lead to
secondary protons or, via charge exchange to neutrons which are then not detected as
CCQE-like any more (\emph{single} proton track).  We find that at tracking detectors the
amount of fake events in the CCQE-like sample is less than at Cherenkov detectors (dashed
and dash-dotted lines almost agree with each other in the lower panel but not in the top
panel).  We conclude that, even if the additional cut on the proton helps to restrict the
background, an error of about 20\% remains since the measured CCQE cross section
underestimates the true one by that amount.  Note that experimental detection thresholds
are not yet taken into account.  Thus, about 20\% of the total cross section has to be
reconstructed by using event generators. In this case, these generators have to be very
realistic in describing the in-medium nucleon-nucleon interactions.
\begin{figure}[tbp]
  \centering
  \includegraphics[scale=\plotscale]{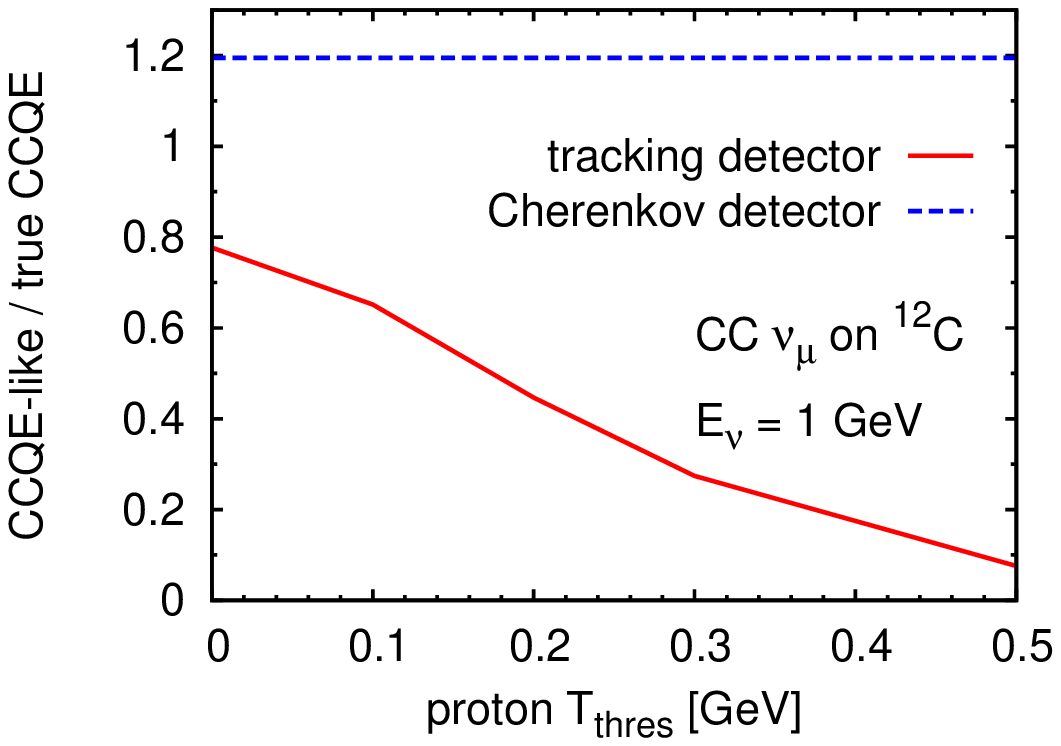}
  \includegraphics[scale=\plotscale]{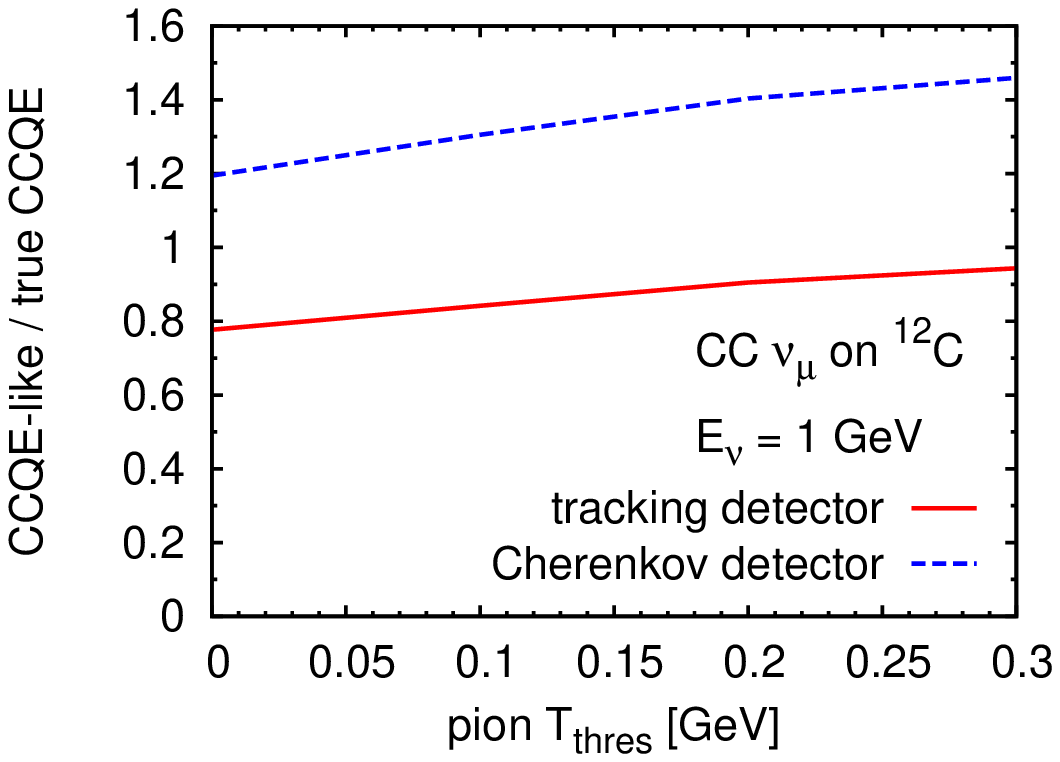}
  \caption{(Color online) Ratio of the CCQE-like to the true CCQE cross section as a function of
  the lower proton (pion) kinetic energy detection threshold for CC $\nu_\mu$ on \carbon{} at
  $E_\nu=1\GeV$. The solid lines are obtained using the tracking detector identification,
  whereas the dashed lines are for Cherenkov detectors.
  \label{fig:CCQElike_over_trueCCQE_thresholds}}
\end{figure}

\begin{figure}[tbp]
  \centering
  \includegraphics[scale=\plotscale]{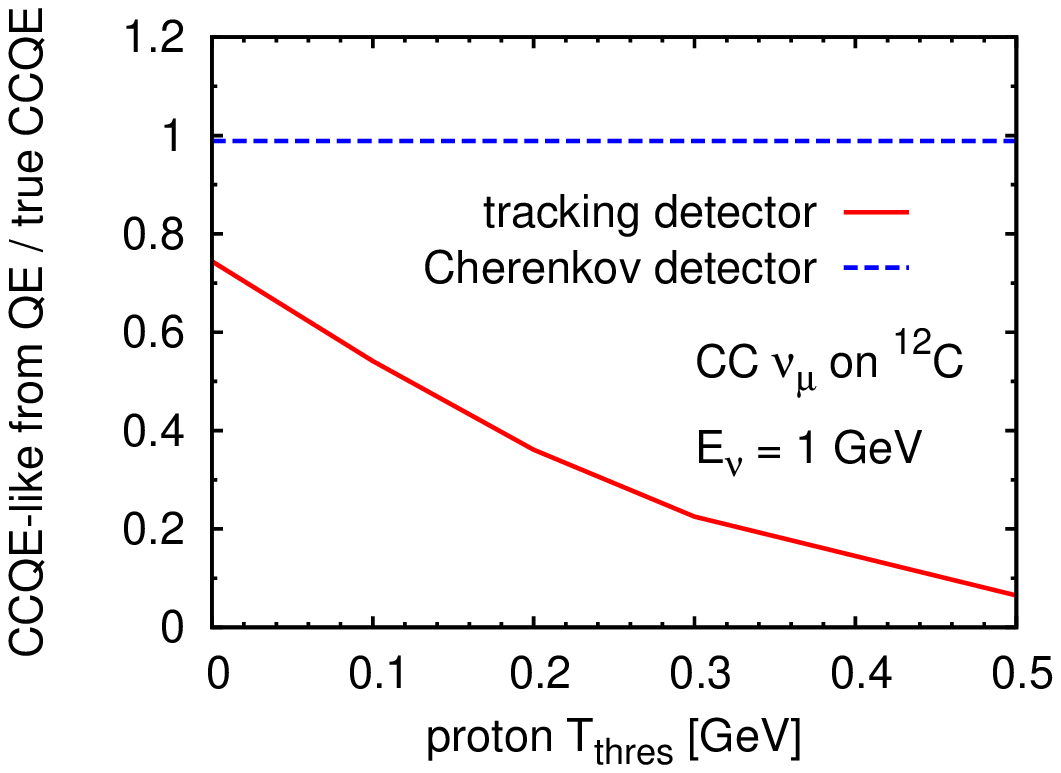}
  \includegraphics[scale=\plotscale]{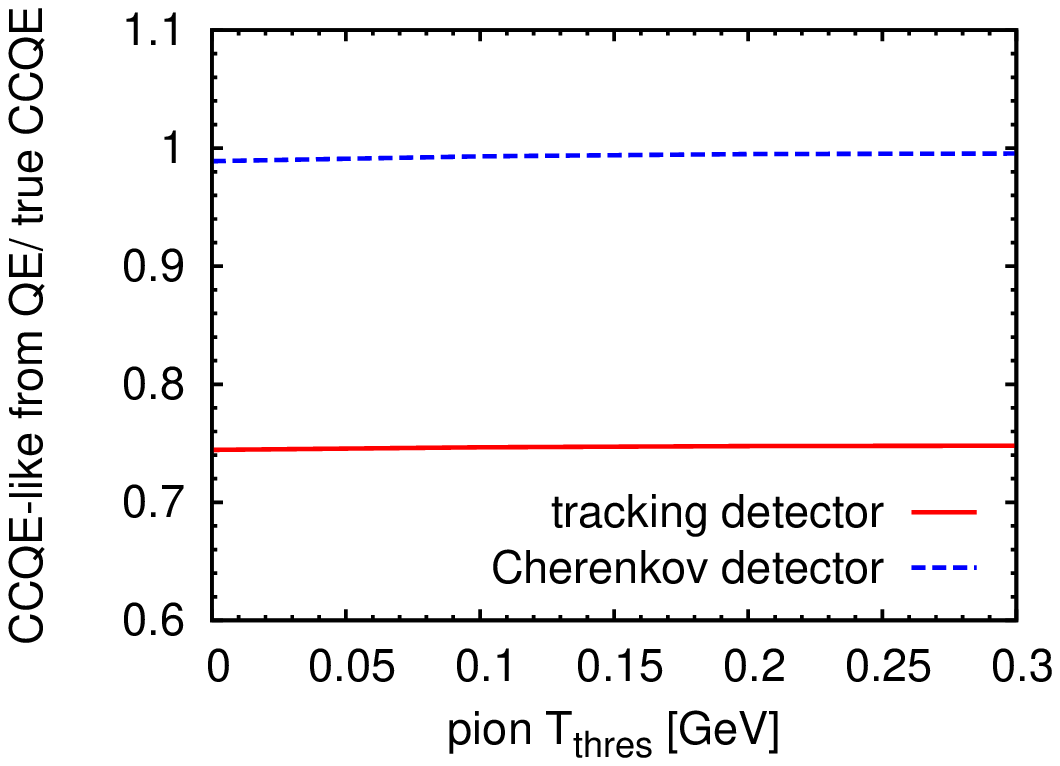}
  \caption{(Color online) Ratio of the QE-induced CCQE-like (i.e., the fraction of the CCQE-like
  event where the initial interaction was QE: $\nu_\mu n \to \mu^- p$) to the true CCQE
  cross section as a function of the lower proton (pion) kinetic energy detection threshold for CC
  $\nu_\mu$ on \carbon{} at $E_\nu=1\GeV$. The solid lines are obtained using the tracking
  detector identification, whereas the dashed lines are for Cherenkov detectors.
  \label{fig:CCQElike_QEinduced_over_trueCCQE_thresholds}}
\end{figure}

To investigate further the relationship between the CCQE-like and true CCQE cross section,
we show their ratio as a function of the lower proton and pion-kinetic energy detection
thresholds in \reffig{fig:CCQElike_over_trueCCQE_thresholds} (see
\refsec{sec:eventselection} for the thresholds applied in present experiments). As the
proton is not at all relevant for the CCQE identification in Cherenkov detectors, the
ratio is independent of the proton kinetic energy detection threshold (dashed line in top
panel). This is very different in tracking detectors which rely on the detected proton ---
here the efficiency is reduced to $\approx$10\% at a proton kinetic energy threshold of
0.5\GeV{} (solid line in top panel). Even at $T_\text{thres}^p=0$, the efficiency does not
exceed 80\% because of charge-exchange processes that lead to the emission of undetected
neutrons and because of secondary proton knockout that leads to multiple-proton tracks.
These effects cause the difference between the solid and the dashed lines in the top panel
of \reffig{fig:QEmethods}.  Focussing on the lower panel of
\reffig{fig:CCQElike_over_trueCCQE_thresholds}, we find that the CCQE-like cross section
increases for both detector types as $T_\text{thres}^\pi$ increases. In this case, even
more events with pions in the final state appear as CCQE-like because then these pions are
below threshold and thus not detected.

\begin{figure}[tbp]
  \centering
  \includegraphics[scale=\plotscale]{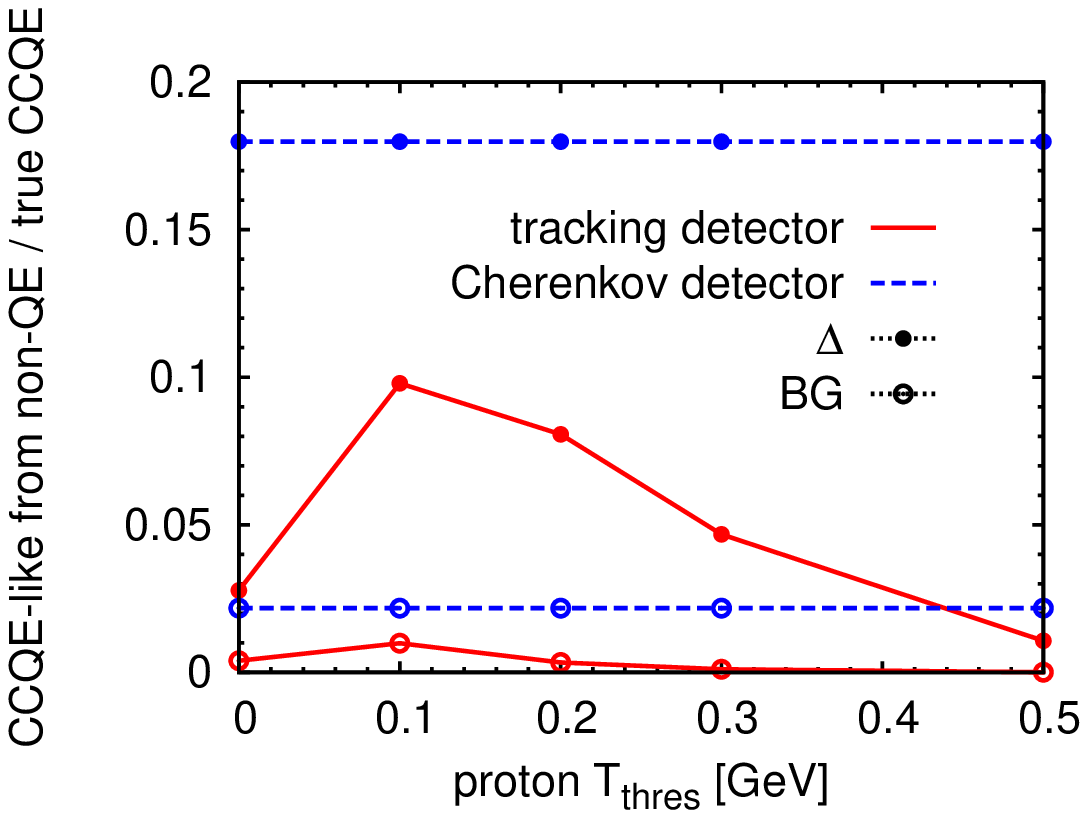}
  \includegraphics[scale=\plotscale]{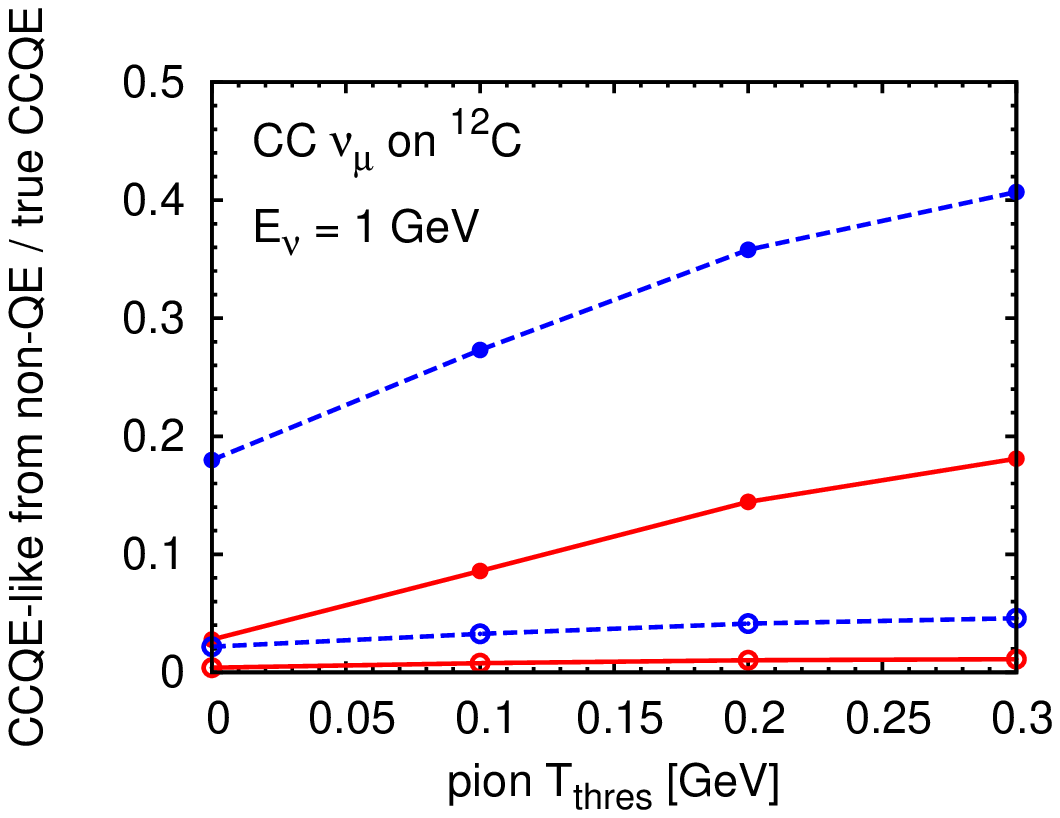}
  \caption{(Color online) Ratio of the \emph{non-QE induced} CCQE-like (i.e., the initial
  interaction was \emph{not} $\nu_\mu n \to \mu^- p$) to the true CCQE cross section as a
  function of the proton (pion) detection kinetic energy threshold for CC $\nu_\mu$ on \carbon{}
  at $E_\nu=1\GeV$. The solid lines are obtained using the tracking detector
  identification, while the dashed lines are for Cherenkov detectors. The bullets
  symbolize the cross sections where the initial reaction is $\Delta$ excitation, whereas
  the open symbols stand for the initial single-pion background. The contribution from
  higher resonances is negligible.
  \label{fig:CCQElike_nonQEinduced_over_trueCCQE_thresholds}}
\end{figure}

The CCQE-like cross section is split into QE and non-QE sources (like $\Delta$ excitation)
in \reffig{fig:CCQElike_QEinduced_over_trueCCQE_thresholds} and
\reffig{fig:CCQElike_nonQEinduced_over_trueCCQE_thresholds}. The top panel of
\reffig{fig:CCQElike_QEinduced_over_trueCCQE_thresholds} shows again the ability of a
Cherenkov-like detector to identify over 98\% of the initial CCQE events (dashed line);
the missing strength is mainly lost into pion channels, that is, the nucleons rescatter
and produce pions such that the event is no longer classified as CCQE-like. This fraction
almost vanishes (the dashed line gets even closer to one in the lower panel) when the pion
kinetic energy threshold increases because then the CCQE-induced pions are no longer
detected and the event counts again as CCQE-like.

Let us now turn to the non-QE CCQE-like cross section displayed in
\reffig{fig:CCQElike_nonQEinduced_over_trueCCQE_thresholds}. Different sources are
indicated: initial $\Delta$ excitation and initial single-pion background reaction (higher
resonances are negligible here and thus not shown). The top panel shows again the
dependence on the proton-kinetic energy threshold, which is not relevant in the Cherenkov
case, where the non-QE CCQE-like contribution adds up constantly to about 18\%. However,
this threshold is important for the tracking detector for the following reason: The non-QE
processes lead not only to single-proton knockout but also to multi-nucleon knockout
through pion absorption processes and rescattering. If the proton threshold is zero, these
processes are not counted because there is more than one proton present. Increasing the
threshold also increases the probability that only one proton is above threshold, in which
case the event is CCQE-like. Above a certain kinetic energy on ($\approx 0.1\GeV$), more
and more protons are below threshold and the ratio decreases again.  The dependence on the
pion kinetic energy threshold is displayed in the lower panel. Here the ratio increases
because, with increasing threshold for the outgoing pion, more and more non-QE events are
misidentified as CCQE.

We note that a realistic muon-kinetic energy threshold of roughly the same magnitude has
no visible influence on the CCQE-like to true CCQE cross-section ratio since the muon
kinetic energy is larger in most cases.

\begin{figure}[tbp]
  \centering
  \includegraphics[scale=\plotscale]{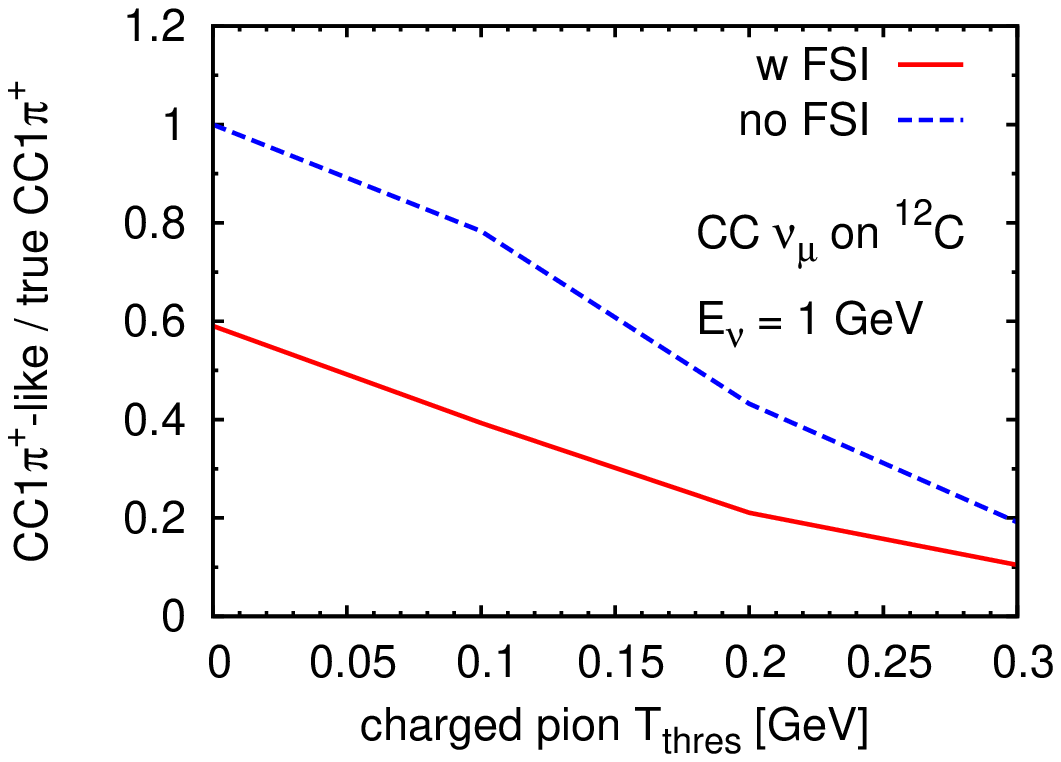}
  \includegraphics[scale=\plotscale]{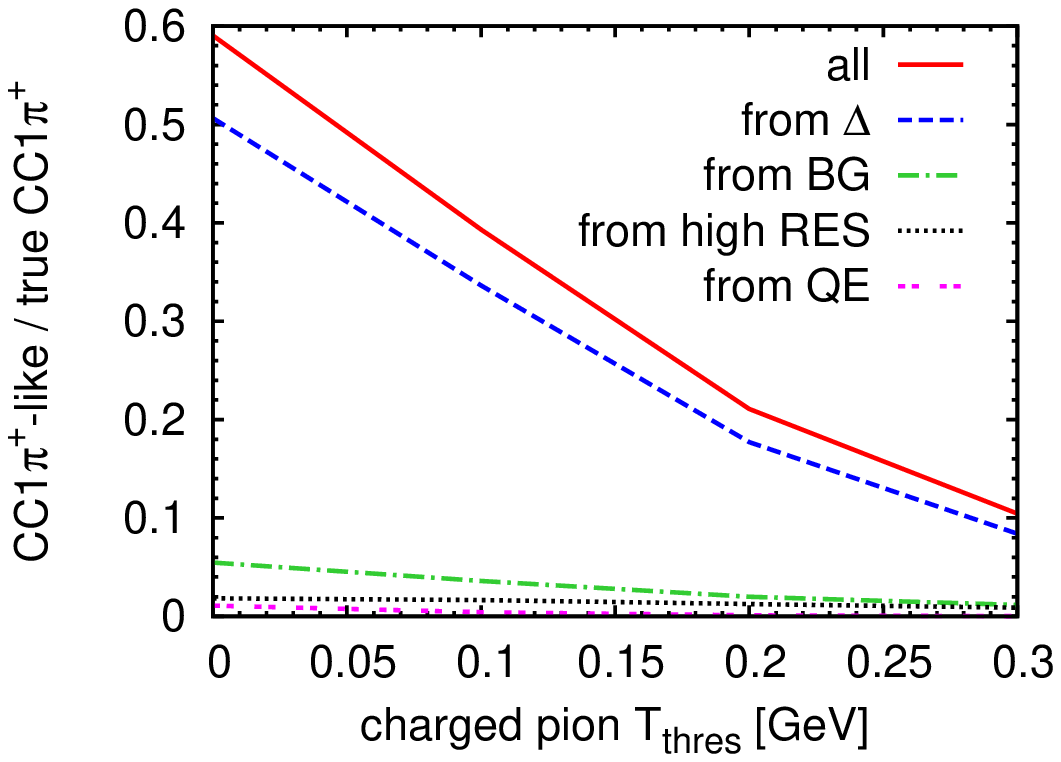}
  \caption{(Color online) Ratio of the CC$1\pi^+$-like to the true CC$1\pi^+$ cross section as a
  function of the pion detection kinetic energy threshold for CC $\nu_\mu$ on \carbon{} at
  $E_\nu=1\GeV$. Top panel: the solid line is obtained with FSI, the dashed without FSI.
  Lower panel: the different contributions to the full result (corresponds to the solid
  line in the top panel) are shown as indicated in the plot.
  \label{fig:CC1pilike_over_trueCC1pi_thresholds}}
\end{figure}

\begin{figure*}[tbp]
  \centering
  \includegraphics[scale=\plotscale]{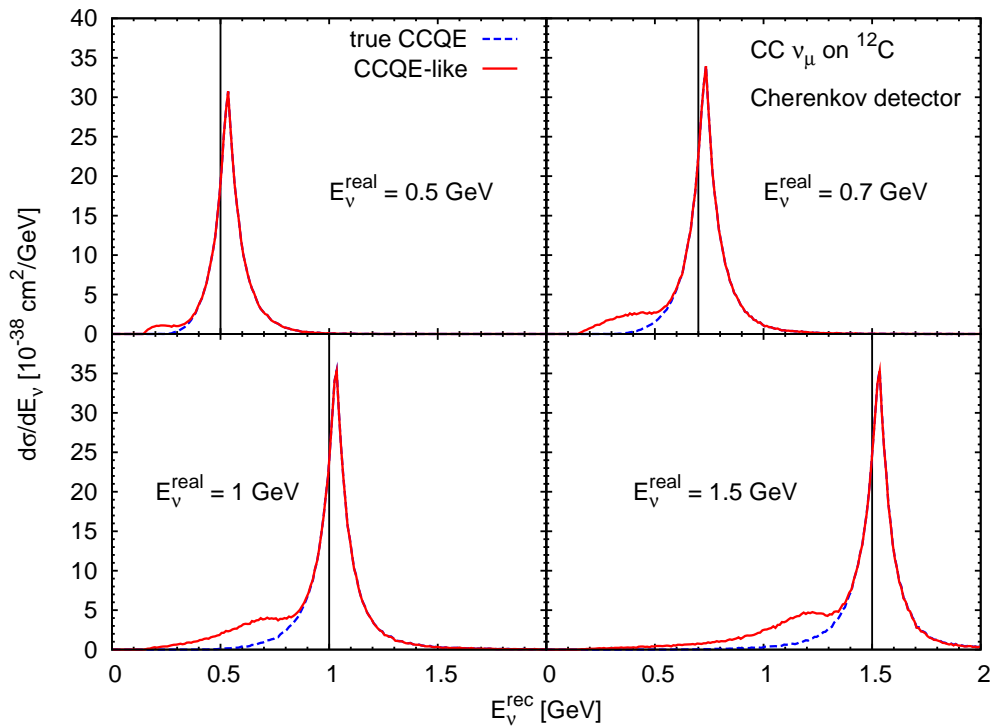}
  \caption{(Color online) Distribution of the reconstructed neutrino energy according to
  \refeq{eq:energy_rec} for $E_\nu^\text{real}=0.5$, $0.7$, $1.0$ and $1.5\GeV$. The
  reconstructed energy denoted by the dashed lines includes only true CCQE events, while
  the solid lines are obtained by reconstructing the energy with CCQE-like events under
  Cherenkov assumptions.
  \label{fig:CC_recEnu_Cherenkov}}
\end{figure*}

\subsection{{CC$1\pi^+$} identification}

The CC$1\pi^+$ reaction is the second largest cross section at the energies of interest in
this work, and the major background to the CCQE signal channel as we have seen in the
previous section.

As in the case of CCQE, the detected CC$1\pi^+$ events can also be masked by FSI. However,
as we will now show, the misidentification is minor and independent of the detector type
--- both of our generic detectors identify CC$1\pi^+$ in the same way. Problematic,
however, is the low efficiency caused by strong pion-absorption effects. The top panel of
\reffig{fig:CC1pilike_over_trueCC1pi_thresholds} shows that already without any threshold
cuts only 60\% of the pions leave the nucleus and can be detected.\footnote{We have
  normalized the true CC$1\pi^+$ to the ``no FSI'' curve at $T^\pi_\text{thres}=0$. Note
  also that we use the data of the Argonne bubble chamber experiment (ANL) as reference
  for our elementary pion production cross section \cite{Leitner:2008ue}.}  Increasing the
pion kinetic energy threshold decreases clearly the CC$1\pi^+$ event rate in the detector.
In the lower panel, we plot the different contributions separately and find that the
$\Delta$ excitation dominates.  Concluding,
\reffig{fig:CC1pilike_over_trueCC1pi_thresholds} shows that the experiment sees only less
than 60\% of all pions, with that number decreasing rapidly with increasing pion kinetic
energy threshold. Therefore, a large part of the total pion yield has to be reconstructed.
Any data on pion production thus contain a major model dependence. This makes it mandatory
to use state-of-the-art and well-tested descriptions of the $\pi N \Delta$ dynamics in
nuclei.


\section{Neutrino energy reconstruction}
\label{sec:energyreconstruction}

\subsection{CCQE}

In long-baseline (LBL) experiments, CCQE events are commonly used to determine the $\nu_\mu$ kinematics.
The neutrino energy has been reconstructed from QE events at the MiniBooNE experiment
\cite{Aguilar:2007ru} using
\begin{equation}
  E_\nu^\text{rec} = \frac{2(M_N - E_B)E_\mu - (E_B^2 - 2M_N E_B + m_\mu^2)} {2\:[(M_N - E_B) - E_\mu + \abskpr \cos\theta_\mu]},   \label{eq:energy_rec}
\end{equation}
with a binding energy correction of $E_B=34\MeV$ and the measured muon energy, $E_\mu$,
and scattering angle, $\theta_\mu$.  The K2K experiment uses the same expression but with
$E_B=0$ \cite{Ahn:2002up}. \refeq{eq:energy_rec} is based on the assumption of quasifree
kinematics on a nucleon at rest.

In \reffig{fig:CC_recEnu_Cherenkov} we plot the distribution of the reconstructed neutrino
energy obtained using \refeq{eq:energy_rec} with $E_B=34\MeV$ for four fixed
$E_\nu^\text{real}$ (0.5, 0.7, 1.0 and 1.5\GeV). The dashed lines show the true CCQE
events only, the solid lines all CCQE-like events (using the Cherenkov definition, but
without any threshold cuts). Both curves show a prominent peak around the real energy
which is slightly shifted to higher $E_\nu^\text{rec}$. This shift is caused by the
difference between our potential and the specific choice of $E_B$.\footnote{See
  \reffig{fig:energyrec_ratio}: the dash-dotted and dotted lines there have been obtained
  with $E_B=0$. We note that in an event simulation the binding energy parameter could be
  adjusted such that the maximum of the reconstructed distribution is at the true energy.
  This is, however, not possible in the actual experiments where the true energy is not
  known.}  The peak has a width of around 100 MeV full width at half maximum (FWHM). This
broadening is entirely caused by the Fermi motion of the nucleons ---
\refeq{eq:energy_rec} assumes nucleons are rest.

While the distribution of the reconstructed energy for the true CCQE events is symmetric
around the peak, this is not the case for the CCQE-like distribution. The reconstruction
procedure now includes also non-CCQE events. However, \refeq{eq:energy_rec} is entirely
based on the muon kinematics and, in the case of $\Delta$-induced non-CCQE events, more
transferred energy is needed than for true CCQE, so the muon energy is smaller.  This
lower muon energy leads then to the second smaller bump at lower reconstructed energies.
Thus, the asymmetry is caused by the non-CCQE events that are identified as CCQE-like.

\begin{figure}[tbp]
  \centering
  \includegraphics[scale=\plotscale]{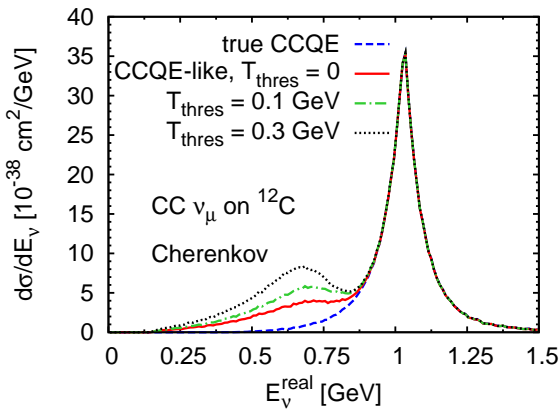}
  \caption{(Color online) Distribution of the reconstructed neutrino energy according to
  \refeq{eq:energy_rec} for $E_\nu^\text{real}=1\GeV$. Using only true CCQE events for the
  reconstruction leads to the dashed line. Including CCQE-like events (Cherenkov
  definition) with various charged pion detection thresholds, one obtains the solid line
  (no pion threshold), the dash-dotted line (100 MeV), and the dotted line (300 MeV).
  \label{fig:CC_recEnu_Cherenkov_pionThres}}
\end{figure}

The asymmetry is very sensitive to detection thresholds, in particular to the kinetic
energy threshold for charged pions (see \refsec{sec:eventselection} for the thresholds
applied in present experiments). We have seen in the previous section that increasing
this threshold also increases the CCQE-like cross section (via the non-CCQE events).
Thus, a higher threshold leads to a more pronounced second bump, as seen in
\reffig{fig:CC_recEnu_Cherenkov_pionThres}.

\begin{figure*}[tbp]
  \centering
  \includegraphics[scale=\plotscale]{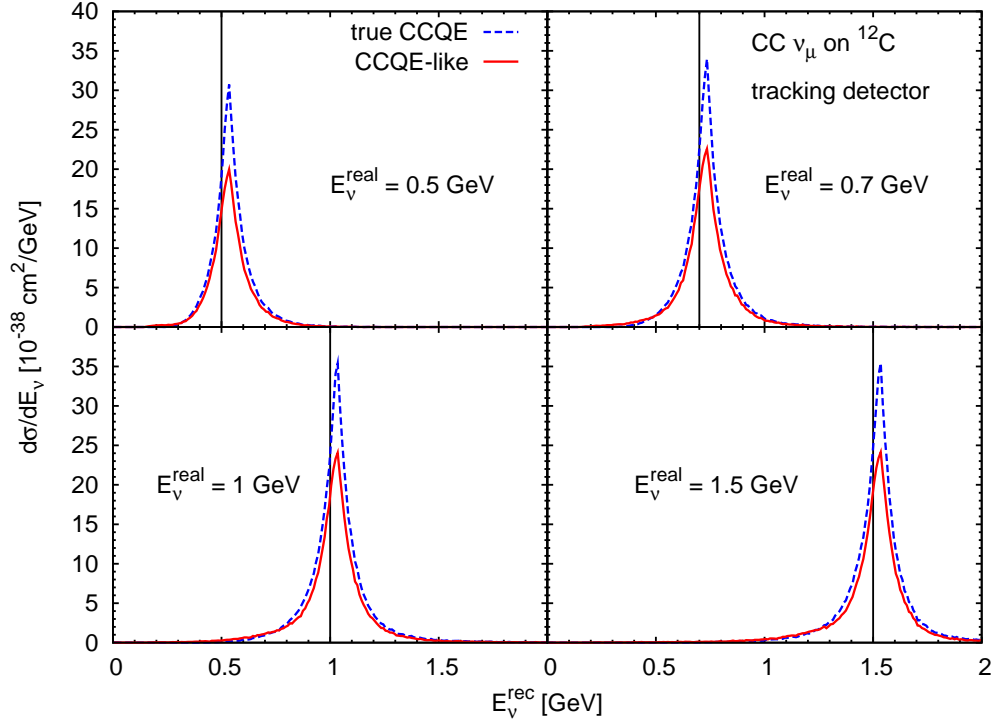}
  \caption{(Color online) Same as \reffig{fig:CC_recEnu_Cherenkov}, but under tracking
    detector assumptions for the CCQE-like events (solid
    lines).\label{fig:CC_recEnu_tracking}}
\end{figure*}

The reconstructed energy under tracking detector assumptions is plotted in
\reffig{fig:CC_recEnu_tracking}. We have seen in the previous section that the tracking
detector allows the extraction of a much cleaner CCQE-like sample than the Cherenkov
detector --- almost no fake, i.e., non-CCQE events spoil the CCQE-like sample.
Consequently, the reconstructed distribution is again symmetric, but at the cost of a
lower detection rate.

\begin{figure*}[tbp]
  \centering
  \includegraphics[scale=\plotscale]{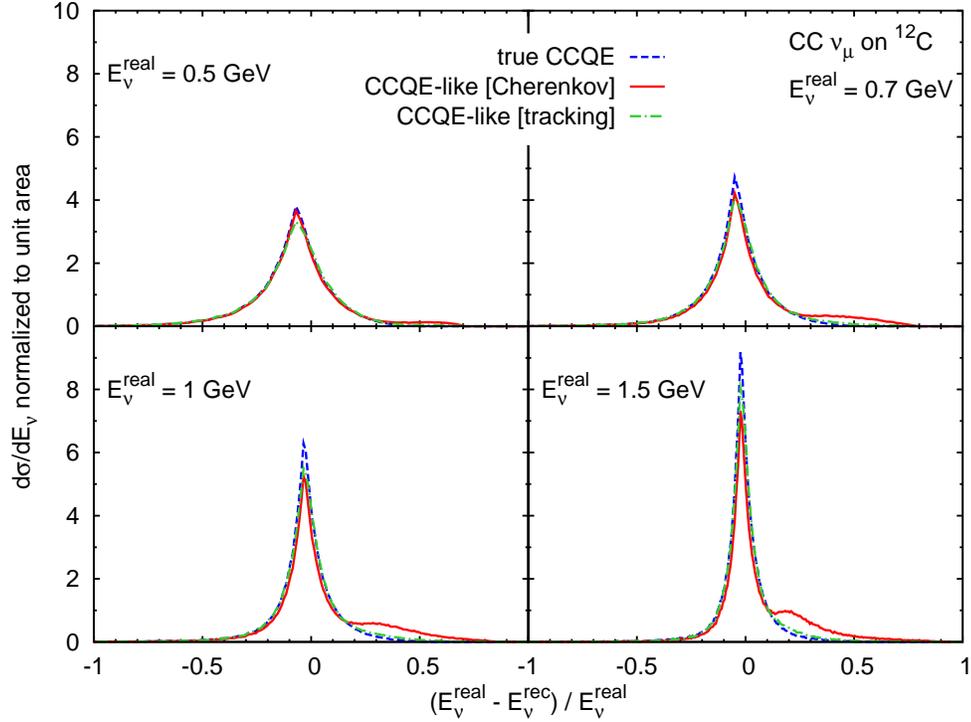}
  \caption{(Color online) Normalized distribution of the reconstructed energy vs.~the relative discrepancy
    using \refeq{eq:energy_rec} for $E_\nu^\text{real}=0.5$, $0.7$, $1.0$, and
    $1.5\GeV$. The dashed lines use true CCQE, the solid ones CCQE-like in a Cherenkov
    detector and the dash-dotted ones CCQE-like events in a tracking detector for the
    reconstruction.
    \label{fig:CC_recEnu_bias}}
\end{figure*}

The previous findings, without any threshold cuts, are summarized in
\reftab{tab:energyrec_expVal_stDev}\footnote{Note that the standard deviations do not
  reflect the low-energy tails caused by the misidentification of events.} and in
\reffig{fig:CC_recEnu_bias}. The former lists the expected values for the reconstructed
energy and the standard deviation, while the latter shows the probability distribution of
the relative discrepancy $(E_\nu^\text{real}-E_\nu^\text{rec})/E_\nu^\text{real}$ for 4
different real energies.  We note that similar investigations by Blondel
\refetal{Blondel:2004cx} and Butkevich \cite{Butkevich:2008ef} result in smaller
discrepancies. Both works consider only CCQE in the initial state, and do not include,
e.g., $\Delta$ excitation.

\begin{table}[tbp]
  \caption[]{Expected value, $E= \int_{0}^{\infty} \dd E_\nu^\text{rec} \; \frac{E_\nu^\text{rec}}{\sigma} \frac{\dd \sigma}{\dd E_\nu}$, and standard deviation, $S= \left(\int_{0}^{\infty} \dd E_\nu^\text{rec} \; \frac{(E_\nu^\text{rec}-E)^2}{\sigma} \frac{\dd \sigma}{\dd E_\nu}\right)^{1/2} $, for the distributions shown in \reffig{fig:CC_recEnu_Cherenkov} and \reffig{fig:CC_recEnu_tracking}. \label{tab:energyrec_expVal_stDev}}
  \centering
  \begin{tabular}{c c c c}
    \hline \hline
    & $E_\nu^\text{real}$ [GeV]  & $E$ [GeV]  & $S$ [GeV] \\
    \hline 
    true CCQE   & 0.5   & 0.55   & 0.09  (17\%) \\
    & 0.7   & 0.74   & 0.12  (16\%) \\
    & 1.0   & 1.03   & 0.15  (15\%) \\
    & 1.5   & 1.52   & 0.16  (11\%) \\ 
    \\
    CCQE-like               & 0.5   & 0.53   & 0.11  (20\%) \\
    (Cherenkov)             & 0.7   & 0.70   & 0.16  (23\%) \\
    & 1.0   & 0.96   & 0.22  (23\%) \\
    & 1.5   & 1.41   & 0.27  (19\%) \\
    \\
    CCQE-like               & 0.5   & 0.54   & 0.10  (18\%) \\
    (tracking)              & 0.7   & 0.73   & 0.13  (18\%) \\
    & 1.0   & 1.02   & 0.17  (16\%) \\
    & 1.5   & 1.50   & 0.19  (13\%) \\
    \hline \hline
\end{tabular}
\end{table}

\begin{figure}[tbp]
  \centering
  \includegraphics[scale=\plotscale]{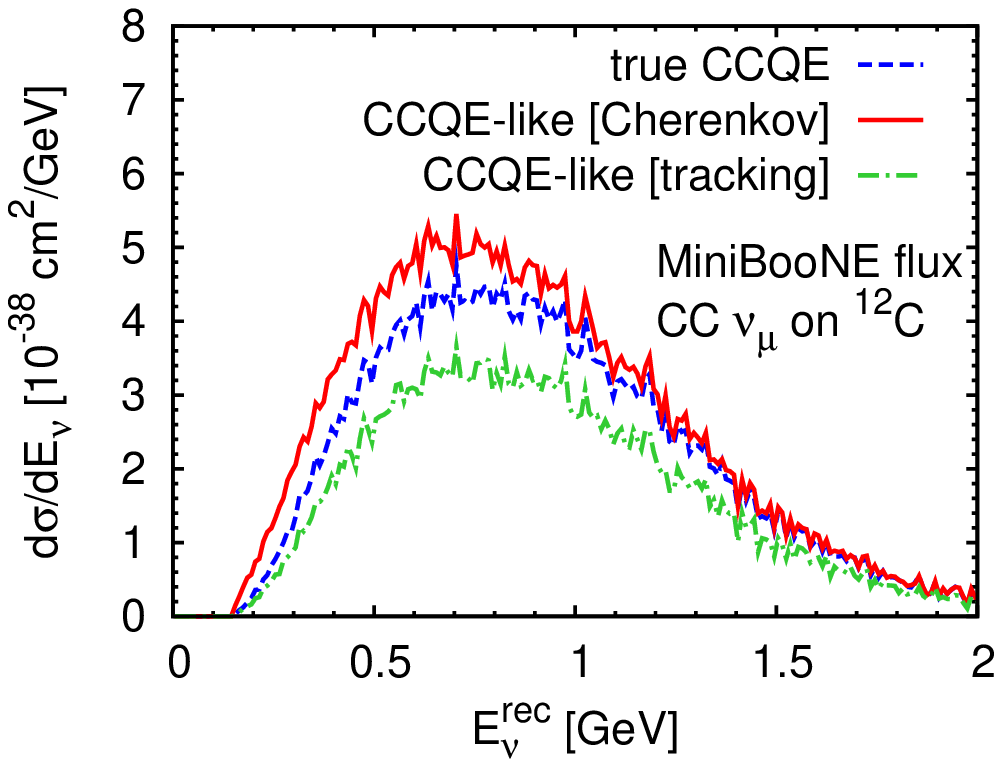}
  \includegraphics[scale=\plotscale]{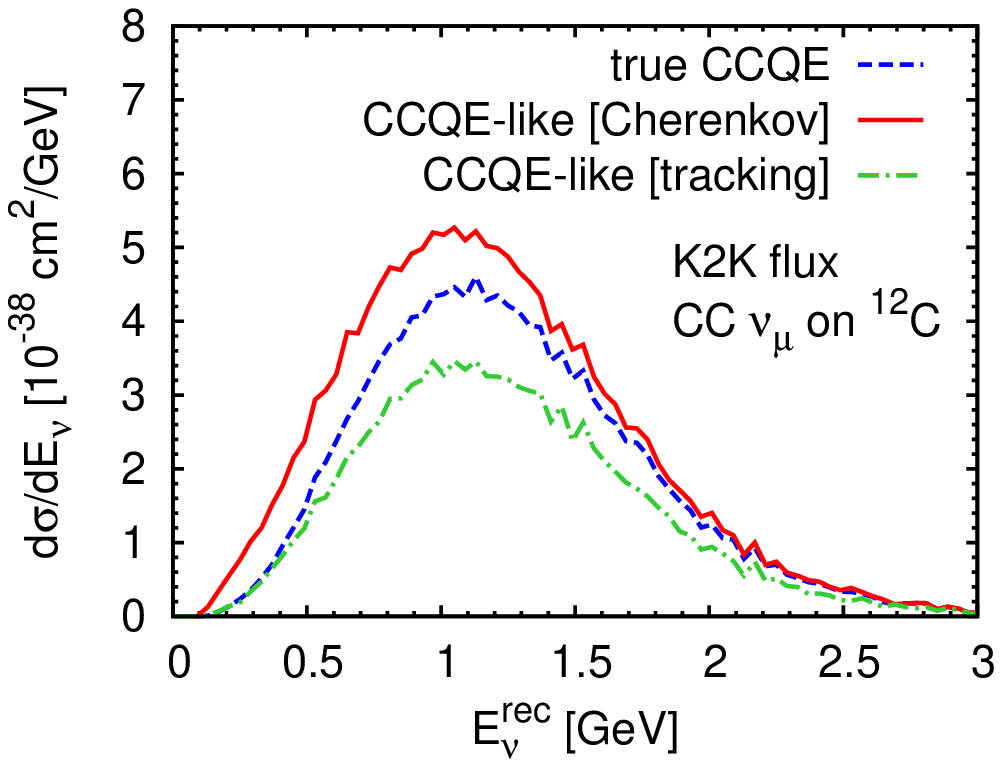}
  \caption{(Color online) Reconstructed energy distribution for the MiniBooNE flux (top panel) and
  K2K flux (lower panel) under different detector assumptions.  \refeq{eq:energy_rec} is
  used for the reconstruction, but with $E_B=0$ in the K2K case.
  \label{fig:CC_flux_recEnu}}
\end{figure}

So far, we have discussed the uncertainties in the energy reconstruction assuming a fixed,
sharp neutrino energy. In reality, the energy distribution of the neutrinos is broad and
thus the question arises how these flux distributions are affected by the reconstruction
procedure. Therefore, we show in \reffig{fig:CC_flux_recEnu} the reconstructed energy
distribution for the MiniBooNE flux (top panel) and the K2K flux (lower panel). Compared
to the true CCQE, we find an enhancement at low reconstructed energies caused by the
non-CCQE induced CCQE-like events in a Cherenkov-like detector (dashed vs.~solid lines,
corresponding to the low-energy bump in \reffig{fig:CC_recEnu_Cherenkov}).  In a tracking
detector, the event rates are reduced (dashed vs.\ dash-dotted lines).

\subsection{CC$1\pi^+$}

The MiniBooNE Collaboration reconstructs the neutrino energy not only using the CCQE
sample, but also using the CC$1\pi^+$ sample.  Based on the observed muon kinematics,
treating the interaction as a two-body collision and assuming that the target nucleon is
at rest inside the nucleus, one finds \cite{AguilarArevalo:2009eb}
\begin{equation}
  E_\nu = \frac{1}{2}\frac{2 M_N E_\mu + M_f^2 - M_N^2 -m_\mu^2}{M_N - E_\mu + \cos\theta_\mu \sqrt{E_\mu^2-m_\mu^2}}, \label{eq:energyrec_ratio}
\end{equation}
where $M_N$ is the mass of the nucleon, $m_\mu$ is the mass of the muon, $\theta_\mu$ its
scattering angle, and $E_\mu$ its energy. $M_f$ is the Breit-Wigner mass of the
\res{P}{33}{1232}.  This formula thus assumes that all pions are produced through the
excitation of the $\Delta$ resonance which is taken to be a state of fixed mass, or, in
other words, its spectral function is taken to be a $\delta$-function. Binding effects are
neglected here. For $M_f=M_N$, this formula agrees with \refeq{eq:energy_rec} for $E_B=0$.

\begin{figure}[tbp]
  \centering
  \includegraphics[scale=\plotscale]{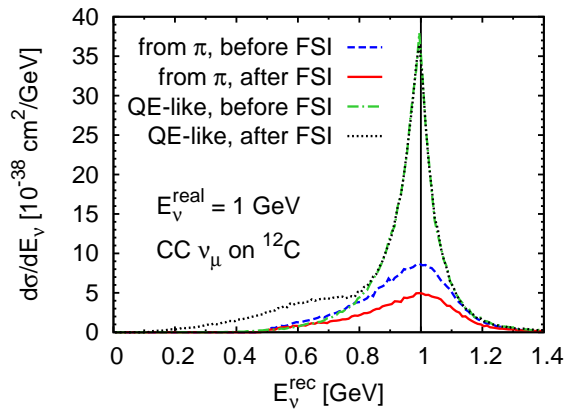}
  \caption{(Color online) Distribution of the reconstructed neutrino energy according to
    \refeq{eq:energyrec_ratio} for $E_\nu^\text{real}=1\GeV$. Shown is the reconstruction
    based on the CCQE-like sample (before and after FSI and Cherenkov assumptions) and based
    on the CC$1\pi^+$ sample (before and after FSI).
    \label{fig:energyrec_ratio}}
\end{figure}

\reffig{fig:energyrec_ratio} shows the reconstructed energy distribution according to
\refeq{eq:energyrec_ratio} for the CCQE-like sample and the CC$1\pi^+$ sample (before and
after FSI). The shape of the dash-dotted and the dotted curves have been discussed before:
Fermi motion broadens the peak and the fake CCQE events cause the bump at lower
reconstructed energies. Also, the reconstructed energy from the pion sample is affected by
Fermi motion (dashed and solid lines). A further broadening comes from the actual shape of
the $\Delta$ resonance which is taken to be of $\delta$ function-like shape in
\refeq{eq:energyrec_ratio}. Overall, the reconstructed energy is centered around the true
energy for both samples, although with a slight tendency to lower reconstructed energies.
\reftab{tab:energyrec_expVal_stDev_pion} lists the expected values for the reconstructed
energy and the standard deviation. Note that the expected value is closer to the real
energy when using the pion sample and that the standard deviation is smaller than in the
CCQE-like case (calculated here also with $E_B = 0$).

\begin{table}[tbp]
  \caption[]{Expected value, $E= \int_{0}^{\infty} \dd E_\nu^\text{rec} \; \frac{E_\nu^\text{rec}}{\sigma} \frac{\dd \sigma}{\dd E_\nu}$, and standard deviation, $S= \left(\int_{0}^{\infty} \dd E_\nu^\text{rec} \; \frac{(E_\nu^\text{rec}-E)^2}{\sigma} \frac{\dd \sigma}{\dd E_\nu}\right)^{1/2}$, for the distributions shown in \reffig{fig:energyrec_ratio} for $E_\nu^\text{real}=1\GeV$. \label{tab:energyrec_expVal_stDev_pion}}
  \centering
  \begin{tabular}{l c c}
    \hline \hline
    & $E$ [GeV] & $S$ [GeV] \\
    \hline 
    from CC$1\pi^+$, before FSI & 0.94  & 0.16  (17\%) \\
    from CC$1\pi^+$, after FSI  & 0.95  & 0.19  (20\%) \\
    CCQE-like, before FSI       & 0.97  & 0.13  (14\%) \\
    CCQE-like, after FSI        & 0.90  & 0.21  (23\%) \\
    \hline \hline
  \end{tabular}
\end{table}

\section{{$Q^2$} reconstruction}

If one assumes a dipole ansatz for the axial form factor, $F_A$, the axial mass, $M_A$, is
the only free parameter in the QE nucleon hadronic current (see, e.g.,
\refcite{Leitner:2008ue} for details; here we use $M_A=1$ GeV).  $M_A$ affects both the
absolute value of the cross section and the shape of the $Q^2$ distribution. Thus, there
are two ways of extracting $M_A$ experimentally (we assume that the vector form factors
are known): (1) $Q^2$-shape-only fit which has the advantage that it does not require
absolute flux normalization, (2) fit to the total cross section. On nuclei, the extraction
of $M_A$ is much more complicated. Nuclear effects change the shape of the $Q^2$
distribution and, consequently, the extracted $M_A$ depends on the model used to relate
measured rates on nuclei to nucleonic form factors. Furthermore, we saw in the previous
section that FSI influence the CCQE identification. Misidentified events are likely to
follow a different $Q^2$ distribution and also affect the total cross section as discussed
in connection with \reffig{fig:QEmethods}.

\begin{figure}[tbp]
  \centering
  \includegraphics[scale=\plotscale]{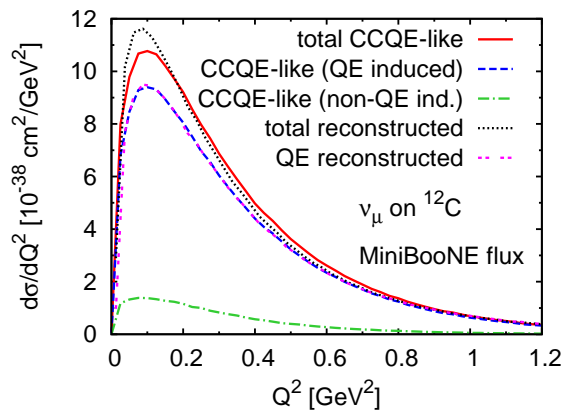}
  \caption{(Color online) Flux averaged $\dsdqs$ distribution of CCQE-like events (solid line) at
    MiniBooNE conditions. The dashed line shows the CCQE-like events induced by CCQE, the
    dash-dotted line shows the non-QE induced CCQE-like contribution. In addition, the
    reconstructed spectra are shown (dotted and double-dashed line). Note that the dashed and the double-dashed lines almost overlap.
    \label{fig:CCQE-like_xsec_reconstruction}}
\end{figure}

Like the neutrino energy, $Q^2$ is not an observable --- it has to be reconstructed from
the measured muon properties. Using \refeq{eq:energy_rec}, we obtain the reconstructed
$Q^2$ via
\begin{equation}
  Q^2 = -m_\mu^2 + 2 E_\nu(E_\mu - \abskpr \cos\theta_\mu) \label{eq:qsrec}
\end{equation}
The neutrino energy itself is reconstructed according to \refeq{eq:energy_rec}, thus
\refeq{eq:qsrec} is also based on the assumption of quasifree kinematics.
\reffig{fig:CCQE-like_xsec_reconstruction} shows the CCQE-like $Q^2$ distribution (solid
line) separated into CCQE-induced CCQE-like (dashed line) and fake CCQE (dash-dotted line)
together with the reconstructed cross section.  If the background subtraction is perfect
(i.e., when the true CCQE sample is isolated and only this sample is used to reconstruct
$Q^2$), then the reconstructed spectrum almost reproduces the true spectrum (dashed and
double-dashed line almost coincide).  If background events, namely non-QE induced events,
are also taken into account for the reconstruction (``total reconstructed'') then, for
the extreme case that no background at all is subtracted, we find an increase at lower
$Q^2$, but then it falls off faster (dotted vs.\ solid line).  The difference is caused by
the different muon kinematics of the ``fake'' events. To conclude, we find that the
reconstruction with the simplified formulas above turns out to be almost perfect when only
true CCQE events are taken into account but not if the whole CCQE-like sample is used to
reconstruct $Q^2$.  The fake events affect both the height and the slope of the $Q^2$
distributions and, thus, the extracted $M_A$ values.

\section{Implications for oscillation parameter measurements in $\nu_\mu$ disappearance
  experiments}

We close this article with a brief discussion on why the exact knowledge of the neutrino
energy is of major importance. The oscillation probability for the transition $\nu_\alpha
\to \nu_\beta$ is given by (within a simplified two-flavor model)
\begin{equation}
  P_\text{osc}(\nu_\alpha \to \nu_\beta) = \sin^2 2 \theta \sin^2 \left( \frac{\Delta m^2 L}{4 E_\nu} \right),
\end{equation}
where $\theta$ is the neutrino mixing angle, $\Delta m^2=m_2^2 - m_1^2$ is the squared
mass difference and $L$ the distance between source and detector.  Consequently,
\begin{equation}
  P_\text{no-osc}=1-P_\text{osc}.
\end{equation}
The oscillation probability depends directly on the neutrino energy $E_\nu$, so measuring
the oscillation parameters $\theta$ and $\Delta m^2$ requires the knowledge of both $L$
and $E_\nu$. In LBL experiments, $L$ and $E_\nu$ are typically chosen such that the
detector is placed in the oscillation maximum or minimum. Commonly, disappearance
experiments measure the neutrino flux at the far detector and compare it to the one
measured at the near detector.\footnote{This is in contrast to neutrino-appearance
  experiments which measure directly the appearance of a different neutrino flavor in the
  beam.} From the difference between both spectra one can determine the oscillation
parameters (compare, e.g., the oscillation analysis performed at the K2K experiment
\cite{Ahn:2006zza}).

\begin{figure}[tbp]
  \centering
  \includegraphics[scale=\plotscale]{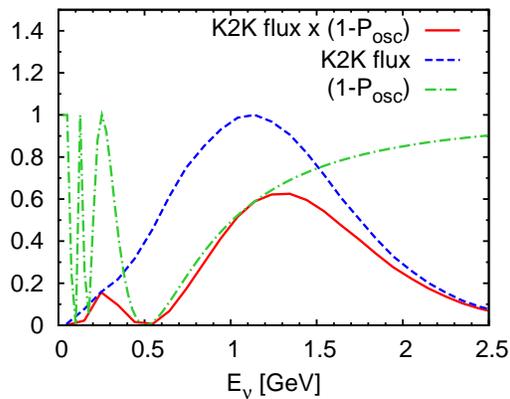}
  \caption{(Color online) Sketch of the influence of neutrino oscillations on the neutrino flux
  (details are given in the text).
  \label{fig:oscillated_flux}}
\end{figure}

A schematic example is given in \reffig{fig:oscillated_flux} for $\theta \approx 45^\circ$
and $\Delta m^2=2.5 \times 10^{-3}$ eV$^2$ (i.e., the parameters measured in $\nu_\mu$
disappearance \cite{Amsler:2008zzb}). It shows the K2K flux with $L=250$ km.  The
un-oscillated spectrum is given by the dashed line, the survival probability by the
dash-dotted line, and its convolution by the solid line.  An exact reconstruction of the
neutrino energy is thus necessary to resolve the oscillated flux, in particular the
characteristic oscillation dip.

\begin{figure}[tbp]
  \centering
  \includegraphics[scale=\plotscale]{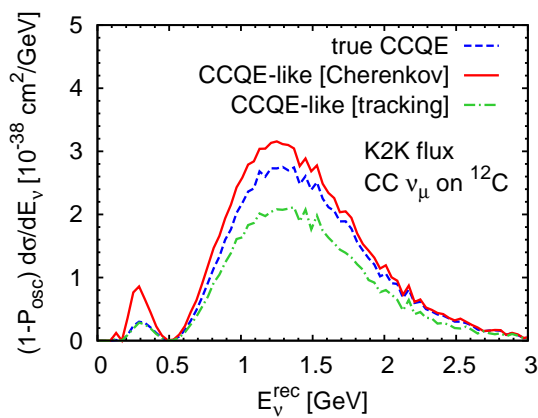}
  \caption{(Color online) Reconstructed energy distribution for the K2K flux folded with
    $1-P_\text{osc}$ under different detector assumptions (compare to
    \reffig{fig:CC_flux_recEnu} where the corresponding un-oscillated spectra are
    shown).
    \label{fig:K2K_osci_spectra}}
\end{figure}

For further illustration, we show in \reffig{fig:K2K_osci_spectra} the convolution of the
oscillation probability with the reconstructed K2K energy flux for the three detection
scenarios introduced before: reconstruction using true CCQE, CCQE-like (Cherenkov), or
CCQE-like (tracking detector) events. Clearly visible is the difference between the
CCQE-like Cherenkov-based reconstruction and the two other methods at low neutrino
energies around the oscillation minimum. Note that, in addition, also the reconstructed
energy is uncertain by up to 20\% (see discussion in \refsec{sec:energyreconstruction}).
This effect is important for $\nu_\mu$ disappearance searches because it affects the
extraction of the oscillation parameters obtained from a fit to these distributions (see
also \refcite{Ahn:2006zza}).

It is even more relevant when an experiment uses different detector types for the near and
far ones, e.g., the K2K experiment uses a tracking detector to measure the un-oscillated
flux and a Cherenkov detector for the oscillated one. Then one has to extract the
oscillation parameters from that difference --- e.g., by comparing the dash-dotted line of
\reffig{fig:CC_flux_recEnu} to the solid line of \reffig{fig:K2K_osci_spectra}. A good
understanding of the energy reconstruction is thus necessary to extract meaningful
oscillation results.

\vspace*{0.7cm}
\section{Summary}

We have applied the GiBUU model to questions relevant to oscillation experiments. This
model provides a theory-based and consistent treatment of in-medium modifications and
final-state interactions. In particular, it describes electron- and photon-induced
reactions successfully which serve as a benchmark for the neutrino-induced processes
discussed here.

The present work addresses the relevance of CC reactions for neutrino-disappearance
experiments. We have argued that a correct identification of CCQE events is relevant
for the neutrino energy reconstruction and, thus, for the oscillation result. A significant
part of CC$1\pi^+$ events is detected as CCQE-like, which is mainly caused by the pion
absorption in the nucleus. We have found that present-day experiments miss the total QE
cross section by about 20\% and the total pion yield by about 40\%. These errors have to
be corrected for by means of event generators so that the final experimental cross
sections contain a significant model dependence. Furthermore, we have investigated the
influence of these in-medium effects on the neutrino energy reconstruction and on the CCQE
cross section, which is the signal channel in oscillation experiments stressing the effect
of final-state interactions.

We conclude that any model that aims to describe the experimental measurements must ---
because of the close entanglement of CCQE and CC$1\pi^+$ on nuclei --- describe both
equally precise, and, in particular, the directly observable rates for nucleon knockout and
$1\pi^+$ production.

\begin{acknowledgments}
  We acknowledge useful discussions with Luis Alvarez-Ruso and the GiBUU team. This work
  has been supported by the Deutsche Forschungsgemeinschaft.
\end{acknowledgments}


\begin{thebibliography}{10}

\bibitem{HARPWebsite}
{HARP},
\newblock \url{http://harp.web.cern.ch/harp/}.

\bibitem{AlvarezRuso:2009mn}
L.~Alvarez-Ruso,
\newblock AIP Conf. Proc. {\bf 1222}, 42 (2010).

\bibitem{gibuu}
Gi{BUU},
\newblock \url{http://gibuu.physik.uni-giessen.de/GiBUU}.

\bibitem{Ankowski:2010yh}
A.~M. Ankowski, O.~Benhar and N.~Farina,
\newblock arXiv:1001.0481.

\bibitem{Martini:2009uj}
M.~Martini, M.~Ericson, G.~Chanfray and J.~Marteau,
\newblock Phys. Rev. C {\bf 80}, 065501 (2009).

\bibitem{Krusche:2004uw}
B.~Krusche {\em et~al.},
\newblock Eur. Phys. J. A {\bf 22}, 277 (2004).

\bibitem{Leitner:2008ue}
T.~Leitner, O.~Buss, L.~Alvarez-Ruso and U.~Mosel,
\newblock Phys. Rev. C {\bf 79}, 034601 (2009).

\bibitem{Leitner:2006ww}
T.~Leitner, L.~Alvarez-Ruso and U.~Mosel,
\newblock Phys. Rev. C {\bf 73}, 065502 (2006).

\bibitem{Leitner:2008wx}
T.~Leitner, O.~Buss, U.~Mosel and L.~Alvarez-Ruso,
\newblock Phys. Rev. C {\bf 79}, 038501 (2009).

\bibitem{AguilarArevalo:2008qa}
MiniBooNE, A.~A. Aguilar-Arevalo {\em et~al.},
\newblock Nucl. Instrum. Meth. A {\bf 599}, 28 (2009).

\bibitem{Gran:2006jn}
K2K, R.~Gran {\em et~al.},
\newblock Phys. Rev. D {\bf 74}, 052002 (2006).

\bibitem{Aguilar:2007ru}
MiniBooNE, A.~A. Aguilar-Arevalo {\em et~al.},
\newblock Phys. Rev. Lett. {\bf 100}, 032301 (2008).

\bibitem{Ahn:2002up}
K2K, M.~H. Ahn {\em et~al.},
\newblock Phys. Rev. Lett. {\bf 90}, 041801 (2003).

\bibitem{Blondel:2004cx}
A.~Blondel, M.~Campanelli and M.~Fechner,
\newblock Nucl. Instrum. Meth. A {\bf 535}, 665 (2004).

\bibitem{Butkevich:2008ef}
A.~V. Butkevich,
\newblock Phys. Rev. C {\bf 78}, 015501 (2008).

\bibitem{AguilarArevalo:2009eb}
MiniBooNE, A.~A. Aguilar-Arevalo {\em et~al.},
\newblock Phys. Rev. Lett. {\bf 103}, 081801 (2009).

\bibitem{Ahn:2006zza}
K2K, M.~H. Ahn {\em et~al.},
\newblock Phys. Rev. D {\bf 74}, 072003 (2006).

\bibitem{Amsler:2008zzb}
Particle Data Group, C.~Amsler {\em et~al.},
\newblock Phys. Lett. B {\bf 667}, 1 (2008).

\end{thebibliography}
\end{document}